\documentclass[preprint, final]{aastex}
\slugcomment{Accepted by {\it The Astrophysical Journal}}
\usepackage{natbib}
\begin{document}
\title{A Test of Pre--Main-Sequence Lithium Depletion Models}
\author{Jennifer C. Yee\altaffilmark{1}}
\affil{Department of Astronomy, Ohio State University}
\author{Eric L. N. Jensen\altaffilmark{2}}
\affil{Department of Physics \& Astronomy, Swarthmore College}

\altaffiltext{1}{E-mail: jyee@astronomy.ohio-state.edu}
\altaffiltext{2}{E-mail: ejensen1@swarthmore.edu}

\newcommand{\teff}{\mbox{$T_{{\rm eff}}$}}
\newcommand{\lbol}{\mbox{$L_{{\rm Bol}}$}}

\begin{abstract}
  Despite the extensive study of lithium depletion during
  pre--main-sequence contraction, studies of individual stars show
  discrepancies between ages determined from the HR diagram and ages
  determined from lithium depletion \citep{Song02, White05} indicating
  open questions in the pre--main-sequence evolutionary models. To
  further test these models, we present high resolution spectra for
  members of the $\beta$ Pictoris Moving Group (BPMG), which is young
  and nearby. We measure equivalent widths of the 6707.8\AA\,
  \ion{Li}{1} line in these stars and use them to determine lithium
  abundances. We combine the lithium abundance with the predictions of
  pre--main-sequence evolutionary models in order to calculate a
  lithium depletion age for each star.  We compare this age to the age
  predicted by the HR diagram of the same model. We find that the
  evolutionary models under-predict the amount of lithium depletion
  for the BPMG given its nominal HR diagram age of $\sim$12 Myr
  \citep{Zuckerman01}, particularly for the mid-M stars, which have no
  observable \ion{Li}{1} line. This results in systematically older
  ages calculated from lithium depletion isochrones than from the HR
  diagram. We suggest that this discrepancy may be related to the
  discrepancy between measured M-dwarf radii and the smaller radii
  predicted by evolutionary models.
\end{abstract}


\section{Introduction}
Lithium depletion during pre--main-sequence (PMS) contraction has been
extensively studied. Studies of open clusters have shown that lithium
depletion is a strong function of both age and stellar mass
\citep[e.g.][]{Soderblom93iii,Barrado04,Mentuch08}. These studies
clearly show that the presence of a strong \ion{Li}{1} line at
$6707.8$\AA\, is an indicator of youth in late-type stars. The
regularity of lithium depletion in open clusters implies that the
lithium line can act as a mass-dependent clock \citep{Jeffries01}. By
combining a measurement of the lithium abundance of a star with
theoretical evolutionary models, one might be able to derive an age of
a PMS star whose distance is unknown, e.g. a young field star. In
order to do this, we need to understand PMS lithium depletion very
well.

Most of the work comparing lithium depletion to HR diagram ages has
been done in open clusters, focusing on comparing the cluster age
derived from the lithium depletion boundary (LDB)\footnote{For stars
that are fully convective before they reach the main sequence ($M<0.3
M_{\odot}$), at a given age there is a sharp boundary between stars
that have fully depleted their lithium and slightly less massive stars
with no evidence of depletion \citep{Basri96,Bildsten97}.} in the
lowest-mass stars with the HR diagram age found from fitting the upper
main sequence \citep[e.g.][]{Barrado99, Stauffer99, Burke04,
Jeffries05,Manzi08}.  Most studies find that the LDB age tends to be
older than the age calculated from upper-main sequence fitting, though
the discrepancy may be smaller for younger clusters \citep{Jeffries05,
Manzi08}. These results are used to argue for the inclusion of
convective overshooting in evolutionary models, since it increases the
ages derived from upper-main sequence fitting without affecting the
LDB age. These older upper-main sequence fitting ages tend to be in
better agreement with LDB ages \citep{Burke04}.

A few studies have been done of individual stars, showing that ages
derived from comparison of lithium abundances with models of lithium
depletion are persistently older than ages derived from the HR
diagram. \citet{Song02} find that the age of the binary HIP 112312
derived from its lithium depletion is $>20$ Myr. They find that this
system is likely a member of the $\beta$ Pictoris Moving Group (BPMG),
whose proposed age derived from pre--main-sequence isochrones on the
HR diagram is $\sim12$ Myr \citep{Zuckerman01}. Likewise,
\citet{White05} find that the age of St 34 derived from its position
on the HR diagram is $8\pm3$ Myr whereas its lithium depletion implies an age
of $>35$ Myr. These results indicate some open questions about lithium
depletion in the evolutionary models.

In order to avoid confusing lithium depletion effects with
temperature- or mass-dependent systematic trends in the models, we
would like to compare HR diagram ages with lithium depletion ages for
individual stars as was done for HIP 112312 and St 34, rather than
comparing the ages of more massive stars determined from one technique
with the ages of less massive stars determined from another. To do
this, we need a coeval group of late-type stars so we can study
lithium depletion over a range of masses. We want the stars to be of
intermediate age, $\sim5$--$80$ Myr; much younger, and very little
lithium depletion has occurred; much older, and they are on the main
sequence, where the HR diagram age is degenerate. We also would like
them to be nearby, so we can measure accurate distances and
luminosities, and bright, so we can get high S/N spectra.

In this paper, we examine lithium depletion in some of the later-type
members of the BPMG, which is ideal for studying lithium depletion
because it is young and nearby ($\sim12$ Myr, 10--50 pc;
\citealt{Zuckerman01}). We compare the lithium depletion of our sample
to the predictions of different theoretical PMS models. We also
calculate ages from these models for individual stars and compare the
HR diagram age of each star with the age implied by its lithium
depletion to see if each model is internally consistent in the two
ages determined for a given star. We begin by presenting our spectra
and lithium equivalent widths in $\S$\ref{sec:data}. We then determine
effective temperatures, luminosities, and lithium abundances for the
stars in our sample in $\S$\ref{sec:analysis}. In
$\S$\ref{sec:discussion}, we compare these data with the HR diagram
and lithium depletion isochrones of theoretical models. We find that
the models systematically under-predict the observed lithium depletion
for M stars (or equivalently, over-predict their lithium depletion
ages compared to HR diagram ages). We suggest that the observed
tendency for models of main-sequence M stars to under-predict stellar
radii may hold the key to resolving the discrepancy between the
observed and predicted lithium depletion in late-type stars. We
conclude in $\S$\ref{sec:conclusions}.

\section{Observations and Data}
\label{sec:data}
We took spectra of ten of the late K and M stars or binary systems
listed in Table 1 of \citet{Zuckerman01}, with separate spectra for
each of the components of the binary stars GJ 799 and BD
$-17^\circ$6128; the spectra are shown in Figures \ref{fig:spec} and
\ref{fig:cd-641208}. We also took a spectrum of a somewhat earlier-type
member of the BPMG, the spectroscopic binary HD 155555 A/B. The
spectra were taken with the 4-meter Blanco telescope at CTIO on six
consecutive nights between 2001 Oct.\ 30 and 2001 Nov.\ 5, using the
echelle spectrograph. The spectrograph covered a wavelength range from
$4800 $ \AA\ to $8400 $ \AA\ with a measured spectral resolving power of
$R \sim 40,000$. The spectra were reduced in IRAF\footnote{IRAF is
  distributed by the National Optical Astronomy Observatories, which
  are operated by the Association of Universities for Research in
  Astronomy, Inc., under cooperative agreement with the National
  Science Foundation.}  according to standard procedures. We flux
calibrated the spectra with respect to two flux standards, HR 9087 and
HR 1544, using {\it standard}, {\it sensfunc}, and {\it calibrate} in
IRAF.  We visually examined the flux calibrated spectra to ensure that
the calibration was consistent between orders and between the two
calibration standards.

\subsection{Spectral Type}
\label{sec:st}
To determine spectral types for our sample, we compared the strength
of the TiO5 band to the strength of the nearby continuum following the
method given in \citet{Reid95}. We calculated the ratio of the flux
from $7126 $\AA--$7135 $\AA\, to the flux from $7042 $\AA--$7046 $\AA.
Using the linear fit relation from \citet{Reid95}, we took the ratio
of these fluxes to determine the spectral types of our sample stars.
According to \citet{Reid95}, this method gives an uncertainty of $\pm
0.5$ spectral subtype. Note that this relation is not well-defined for
spectral types much earlier than M0 because the TiO bandheads are very
weak or disappear entirely. Thus, the spectral types measured for HIP
29964, BD $-17^{\circ}$6128 A, and CD $-64^{\circ}$1208 are more
uncertain, although probably correct in a relative sense.
Additionally, since the relation gives a numeric value for the
spectral type, with 0 representing M0, it is not clear how to assign
that number to the Morgan-Keenan system for values $\leq -1$ since
some authors argue that K6 is not a distinct spectral type
\citep[e.g.][]{Kirkpatrick91}.  We follow the convention of omitting
K6 as a spectral type and assign $-1$ to K7 and $-2$ to K5.  Thus, we
give the spectral types of HIP 29964 and BD $-17^{\circ}$6128 A as
K5.4 and K5.7, respectively, corresponding to $-1.6$ and $-1.3$ as
measured from the relation given in \citet{Reid95}.
\citet{Zuckerman01} give the spectral types of these stars as K6/7 and
K7/M0, respectively, and \citet{Torres06} find K4 and K6. The spectral
types of the stars later than K7 agree within the errors with the
types given in \citet{Zuckerman04}, with the exception of HD 155555 C,
for which \citet{Zuckerman04} give M4.5.  However, our value of M3.5
for this star is in agreement with the M3 spectral type found by
\citet{Torres06}.  BD $-17^{\circ}$6128 B was not listed independently
in \citet{Zuckerman04}, but an estimate of its spectral type is given
in \citet{Neuhauser02} as M0--2, which is close to the spectral type
of M2.9 that we find here.

\subsection{Equivalent Width of \ion{Li}{1}}
\begin{figure}
\includegraphics[width=6in]{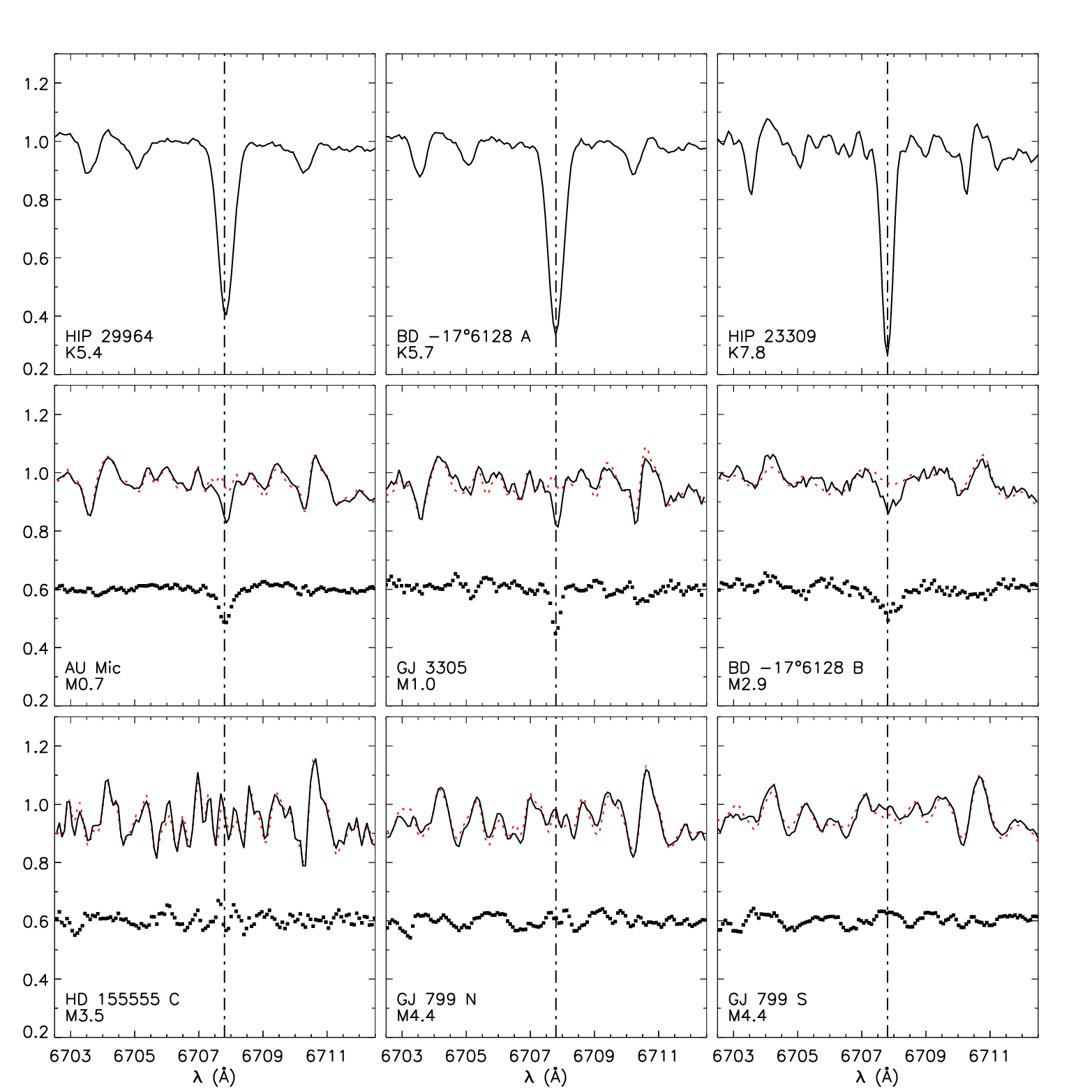}
\caption{Spectra of $\beta$ Pictoris Moving Group members in the
vicinity of the Li line ($\lambda=6707.8$ \AA); the line's position is
marked as a vertical line. The solid curve shows the observed
spectrum. For the M stars, the dotted line (red in the on-line
edition) is the spectrum of a main-sequence template of similar
spectral type that has been artificially broadened and shifted to
produce the best match with the observed spectrum.  Filled squares
show the difference between the broadened template and the observed
spectrum, on the same scale as the observations but with an offset of
0.6 added for clarity. The Li line is clearly detected in the three stars with
spectral types earlier than M3 (middle row), but it is not present in the
later-type stars (bottom row).\label{fig:spec}}
\end{figure}

Using the {\it splot} routine in IRAF, we measured the equivalent
widths of all identifiable \ion{Li}{1} lines at $6707.8 $\AA\,
(EW(\ion{Li}{1}))\footnote{The spectra of M dwarfs are so rich with
  lines that the observed pseudo-continuum is significantly below the
  true continuum level. Because of this, for these stars we actually
  measure a pseudo-equivalent width rather than a true equivalent
  width. For simplicity, we use the term `equivalent width' to refer
  to both equivalent widths (as measured for K and earlier spectral
  types) and pseudo-equivalent widths (as measured for the M
  dwarfs).}. No attempt was made to deblend the \ion{Fe}{1} line at
$6707.4 $\AA\, from the \ion{Li}{1} line. This should not affect our
later analyses of Li abundances since curves of growth for the coolest
stars are often derived from measuring the blended lines in model
spectra, and thus they also include in the quoted EW(Li) values the
contribution of the \ion{Fe}{1} line. The uncertainties given reflect
the difference between choosing the height of the continuum to be the
maxima or the mean of nearby features. The \ion{Li}{1} line is not
detected in HD 155555 C, GJ 799 N, or GJ 799 S. Figure \ref{fig:spec}
shows a comparison of the spectra of the M dwarfs with spectra of
main-sequence stars of similar spectral type that have been
rotationally broadened to match the observed projected rotational
velocity ($v \sin i$). The difference spectra show that while the line
is weak in the early-M stars, it is definitely detected, whereas it is
clearly not present in the mid-M dwarfs. We measure upper limits in
the EW(Li) of 30 m\AA\, for HD 155555 C, 10 m\AA\, for GJ 799 N, and
10 m\AA\, for GJ 799 S.

\begin{deluxetable}{llcrlrllrrl}
\rotate
\tablewidth{0pt}
\tabletypesize{\footnotesize}
\tablecaption{\label{tab:data}Late-Type Members of the $\beta$ Pictoris Moving Group}
\tablehead{
\colhead{}&\colhead{}         & \colhead{Spectral} & \colhead{Distance} & \colhead{V} & \colhead{$v \sin i$} & \colhead{$T_{eff}$}        & \colhead{$L_\star$} & \colhead{EW(\ion{Li}{1})} & \colhead{A(Li)} &\colhead{}\\
\colhead{Name}&\colhead{Other Name}     & \colhead{Type}     & \colhead{(pc)}     & \colhead{(mag)} & \colhead{(km s$^{-1}$)}& \colhead{$(K)$} & \colhead{$(L_{\odot})$} & \colhead{(m\AA\,)} & \colhead{} &\colhead{Notes}\\
}
\tablecolumns{10}

\startdata
HD 155555 A& V824 Ara  & G5 IV& $31.4\pm0.5$ & \phn7.21&34& $5770^{+50}_{-110}$& $1.1\phn\pm\, 0.1\phn$& $\phm{<}120  \pm\, 20$  & \nodata & 1,2,4\\
HD 155555 B& V824 Ara &K0 IV/V&$31.4\pm0.5$ & \phn8.08&33& $5250^{+125}_{-90}$& $0.54 \pm\, 0.06$ & $\phm{<}250  \pm\, 60$  & \nodata &1,2,4\\
HIP 29964& AO Men   & K5.4 & $38.6\pm1.3$ & \phn9.77&15&$4250^{+140}_{-150}$& $0.27 \pm\, 0.04$ & $\phm{<}370  \pm\, 40$ & $ 2.5\pm 0.3$
 &1,2,3\\
BD $-17^{\circ}6128$ A& HD 358623 A &K5.7&$45.7\pm1.6$\tablenotemark{a}&10.6\tablenotemark{b}\phn& 14 & $4130^{+150}_{-130}$& $0.25 \pm\, 0.025$\tablenotemark{c} & $410 \pm\, 20$ & $ 2.6\pm 0.2$
 & 1,2,5\\
CD $-64^{\circ}1208$& \nodata&K7.2&$28.6\pm0.2$&\phn9.54& 110 & $4020^{+140}_{-110}$&$0.21 \pm\, 0.04$& $\phm{<}460 \pm\, 40$ & $ 2.7\pm 0.2$
 & 1,2\\
HIP 23309& CD -57$^{\circ}$1054  & K7.8 & $26.8\pm0.8$ & 10.01 &  8 & $3890^{+100}_{-90}$ & $0.15 \pm\, 0.03$ & $\phm{<}370  \pm\, 50$  & $ 2.1\pm 0.3$
& 1,2,3\\
AU Mic&  GJ 803  & M0.7 &\phn$9.9\pm0.1$&\phn8.81&9& $3750^{+70}_{-70}$  & $0.07 \pm\, 0.01$ & $120 \pm\, 60$ &$-0.1\pm 0.7$
  &1,2,3\\
GJ 3305& \nodata & M1.0 & $29.4\pm0.3$ & 10.59 &  5 & $3710^{+70}_{-70}$  & $0.13 \pm\, 0.02$ & $\phm{<}110  \pm\, 60$ & $-0.4\pm 0.7$
 &3\\
BD $-17^{\circ}6128$ B& HD 358623 B &M2.9&$45.7\pm1.6$\tablenotemark{a}& \nodata & $20$\tablenotemark{d}&$3390^{+70}_{-70}$& $0.05\pm\, 0.01$ & $\phm{<}130 \pm\, 60$ & $-0.2\pm 0.7$
 & 5\\
HD 155555 C&\nodata  & M3.5 & $31.4\pm0.5$\tablenotemark{e} & 12.71 &  6 & $3340^{+70}_{-70}$  & $0.05 \pm\, 0.01$ & $<30$ &  $<-1.4$& 2\\
GJ 799 N& AT Mic    & M4.4 & $10.7\pm0.4$ & 11.02 & 10 & $3210^{+70}_{-70}$  & $0.04 \pm\, 0.01$ & $<10$ &  $<-1.4$& 2,3\\
GJ 799 S& AT Mic    & M4.4 & $10.7\pm0.4$ & 11.02 & 16 & $3210^{+70}_{-70}$  & $0.04 \pm\, 0.02$ & $<10$ &  $<-1.4$ & 2,3\\
\enddata

\tablecomments{Except where noted, distances are from the updated
  Hipparcos catalog \citep{vanLeeuwen07}, V magnitudes are from \citet{Zuckerman01}, and the other data are our own measurements. The given $v \sin i$ is the average of the literature values cited in the last column. (1) $v \sin i$ given in \citet{Zuckerman01}, (2) $v \sin i$ given in \citet{Torres06}, (3) $v \sin i$ given in \citet{Scholz07}, (4) Spectral Type from \citet{Zuckerman01}, (5) Luminosity from \citet{Neuhauser02}}

\tablenotetext{a}{Assuming that BD $-17^{\circ}$6128 A and B have the
  same parallax as their companion HD 199143.}
\tablenotetext{b}{Value for the combined system}
\tablenotetext{c}{We give the corrected uncertainty in the luminosity of BD $-17^{\circ}6128$ A \citep{Neuhauser09}.}
\tablenotetext{d}{The $v \sin i$ of BD $-17^{\circ}6128$ B is not given in the literature. The value given here was estimated from a rotationally broadened stellar template ($\S$\ref{sec:analysis}).}
\tablenotetext{e}{Assuming that HD 155555 C is at the same distance as HD 155555 A/B.}

\end{deluxetable}

\section{Analysis}
\label{sec:analysis}

\subsection{Luminosity and Rotational Velocity}
We determined the luminosity and projected rotational velocity ($v \sin i$) of
each star based on data from the literature. We calculated the
luminosity of each star using $V$ magnitudes from Table 1 in
\citet{Zuckerman01}. We assigned a fixed value of $\sigma_V = \pm 0.1$
as the error on the V magnitudes to account for measurement errors and
variability. We used bolometric corrections from \citet{Kenyon95}
appropriate to the spectral type of the star. Distances to each star
came from the new reduction of the Hipparcos catalog
\citep{vanLeeuwen07}. The $V$ magnitude of BD $-17^{\circ}6128$ is the
magnitude of the combined system, so we adopt the luminosities given
in \citet{Neuhauser02} for the A and B components.

For the $v \sin i$ of each star, we used the average of the
values given in \citet{Zuckerman01}, \citet{Torres06}, and
\citet{Scholz07}. For CD $-64^{\circ}1208$, we used the value given in
\citet{Torres06}, $v \sin i = 110 \,\mathrm{km\,s^{-1}}$, as
\citet{Zuckerman01} simply give $v \sin i \gtrsim 100
\,\mathrm{km\,s^{-1}}$. BD $-17^{\circ}6128$ B does not have a $v \sin
i$ given in the literature. We measured the $v \sin i$ by artificially
broadening a template star spectrum of similar spectral type
(Fig. \ref{fig:spec}). We find that the $v \sin i$ of BD
$-17^{\circ}6128$ B is $20\pm5$ km/s.

\subsection{Effective Temperature}

We converted the measured spectral type for each star into an
effective temperature \teff\, using empirical spectral type-\teff\,
relations to calculate the effective temperature of each star. For
spectral types M1--M9, we used the temperature scale given in
\citet{Luhman99} for ``intermediate'' stars, since our young stars can
be expected to fall in this class. We combined this with the table
given in \citet{Kenyon95} for earlier spectral types and interpolated
to get effective temperatures for our stars. Changes within the
spectral type uncertainties lead to typical changes in effective
temperature of $\pm \sim100$K. These data are given in Table
\ref{tab:data}.

\subsection{Lithium Abundances}
We calculated lithium abundances ($A(\mathrm{Li})=12+\log
(N(\mathrm{Li})/N(\mathrm{H}))$;\,\citealt{Jeffries00}) for our stars
using the curves of growth given in \citet{Soderblom93iii} and
\citet{Palla07} to convert the measured equivalent widths to lithium
abundances, following the method used by \citet{Sestito08}. We assumed
$\log g = 4.5$, which is appropriate for late-type stars descending
onto the main sequence. We linearly interpolated the curves of growth
in $A(\mathrm{Li})$--EW(\ion{Li}{1}) space and used a linear
extrapolation for abundances $<0.0$ dex. For the three stars with
effective temperature 3600K $<\teff<4000$K, we linearly interpolated
between the two tables. For the stars for which we have only an upper
limit to the equivalent width, the calculated abundance is necessarily
also an upper limit. The quoted uncertainties are the result of
individually perturbing both the effective temperatures and
EW(\ion{Li}{1})s within their uncertainties and adding the effects on
the derived abundances in quadrature. The results are given in Table
\ref{tab:data}.

Because HD 155555 A/B is a spectroscopic binary, we were not able to
determine its lithium abundance from the EW(\ion{Li}{1})s. The height
of the continuum is roughly twice what would be measured for each star
individually; therefore, the lithium equivalent widths of the
individual stars would be larger than the equivalent widths given in
Table \ref{tab:data}, which are measured from the combined
spectrum. We use the abundances given in \citet{Randich93} of 3.5 for
HD 155555 A and 3.7 for HD 155555 B. Based on the discussion in that
paper and in \citet{Pasquini91}, we assume an uncertainty of $\pm 0.3$
dex on those abundances.

\subsection{Lithium Depletion Boundary}
We did not detect the LDB of the BPMG. The latest-type stars in our
sample, HD 155555 C, GJ 799 N, and GJ 799 S, showed no observable
lithium lines with very sensitive upper limits
(Fig. \ref{fig:spec}). The LDB of the BPMG has been detected in two
binary systems that are members of this group: HIP 112312 A/B and V343
Nor A/B \citep{Song02,Song03}. The $6707.8$\AA\, Li line was detected in
the lower-mass stars in these binaries whose spectral types are M4.5
and M5, respectively. The stars in our sample with non-detections of
the lithium line have slightly earlier spectral types, which is
consistent with the detected LDB. Comparison of the data with
evolutionary models (see below, Fig. \ref{fig:evolmod}) implies an LDB
age of $\gtrsim 30$ Myr. Our data support the conclusion of
\citet{Song02} that there is a discrepancy between the LDB age of this
group and the proposed age from HR diagram isochrones of
$12^{+8}_{-4}$ Myr \citep{Zuckerman01}.

\subsection{Comparison to Other Work}
To date, two other groups have published lithium abundances and
effective temperatures for some members of the BPMG
\citep{Torres06,Mentuch08}. Given the variation in effective
temperatures and lithium abundances measured by each group, our data
are consistent with previous measurements. Overall, we see the same
trend as other groups with lithium decreasing as a function of
decreasing spectral type and only upper limits in lithium abundance
measured for GJ 799 N and S.

\subsection{Comments on Individual Objects}
\subsubsection{CD $-64^{\circ}1208$}
\begin{figure}
\includegraphics[width=6in]{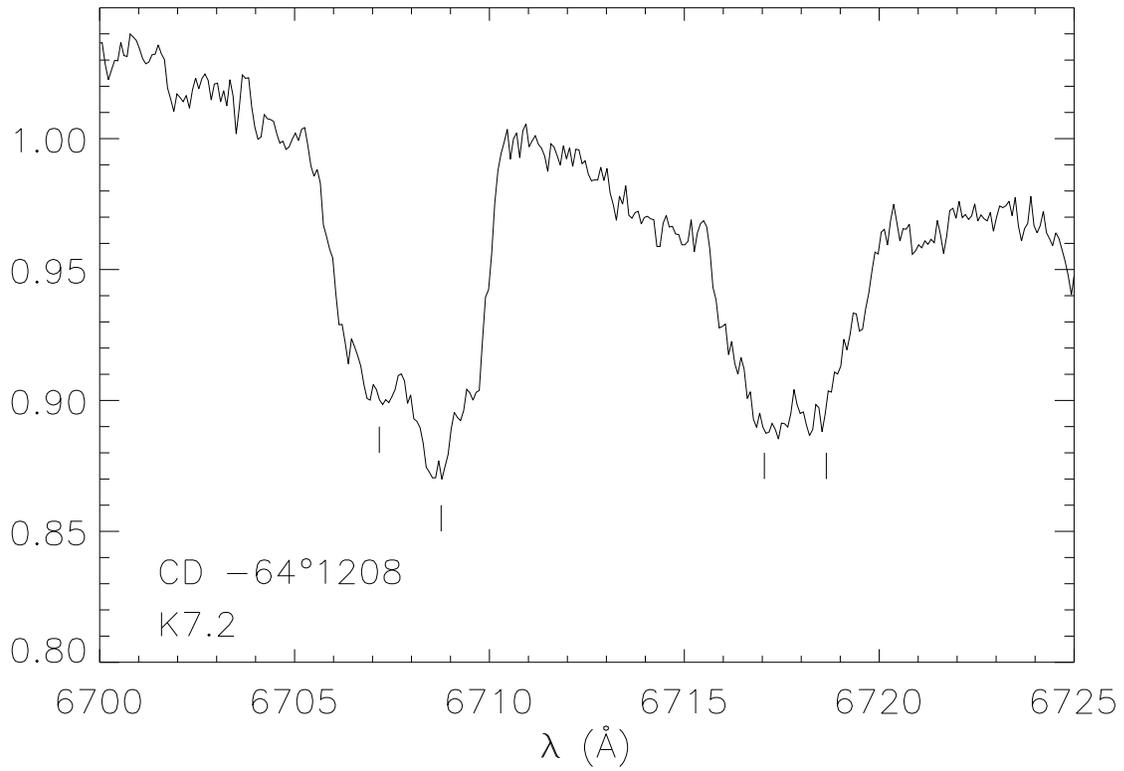}
\caption{Spectrum of CD $-64^{\circ}1208$ showing the \ion{Li}{1}
($6707.8\,$\AA\,) and \ion{Ca}{1} ($6717.7\,$\AA\,) lines. Note that
each line appears to have two components, shown by tick marks at
measured RVs of $-28$ and 43 km/s. \label{fig:cd-641208}}
\end{figure}
Examination of the $6707.8$\AA\, \ion{Li}{1} line and the \ion{Ca}{1}
line at $6717.7$\AA\, suggests that CD $-64^{\circ}1208$ is a
double-lined spectroscopic binary (Fig.\
\ref{fig:cd-641208}). Cross-correlation of the spectrum with radial
velocity standards of known velocity and similar spectral type gives
velocities of $-28\pm8$ km/s for the secondary and $43\pm5$ km/s for
the primary.  Based on the height of the cross-correlation peaks, the
secondary has roughly 70--80\% the luminosity of the primary.  We note
that \citet{Torres06} also flag this system as ``SB?", i.e. a possible
spectroscopic binary.  Our derived spectral type of K7 agrees with
that of \citet{Riaz06} and is similar to the K5 spectral type found by
\citet{Torres06}.  We note that the $v \sin i$ of 110 km/s measured by
\citet{Torres06} is likely an overestimate.  The rotation is rapid,
but some of the line broadening is due to the velocity separation of
the two stars. If the luminosities and effective temperatures of the
two components are similar, as seems likely from the spectrum, the
equivalent widths of the individual lithium line and the lithium
abundances of the individual components should be roughly the same as
those given in Table \ref{tab:data}.

\subsubsection{AU Mic}
If AU Mic is the same age as the other members of the BPMG, it appears
to be under-luminous in the HR diagram (Fig. \ref{fig:evolmod}). We
observe $\log (L/L_{\odot}) = -1.15$, but from the figure we might
expect a value closer to $\log (L/L_{\odot}) = -0.95$ for it to lie on
the same HR diagram isochrone as the other BPMG members. AU Mic is
known to have an edge-on debris disk \citep[cf.][]{Kalas04}, which
could lead to some extinction. Given the difference between our
observed and expected values for the luminosity, if AU Mic is coeval
on the HR diagram, then we estimate an extinction due to the debris
disk of $A_V \sim 0.5$ mag. Since this extinction is not well
characterized, and it is dependent on an assumption of coevality, we
do not take it into account in our analysis. This proposed extinction
should not affect our measurement of the effective temperature since
that is determined from the spectral type which is measured over a
small range in wavelength. However, it does affect any age determined
from the HR diagram position. If the luminosity we observe is less
than the intrinsic luminosity of AU Mic, we would overestimate the age
of the star (see Fig. \ref{fig:evolmod}).

\section{Comparison to Evolutionary Models}
\label{sec:discussion}

\begin{figure}
\includegraphics[width=6in]{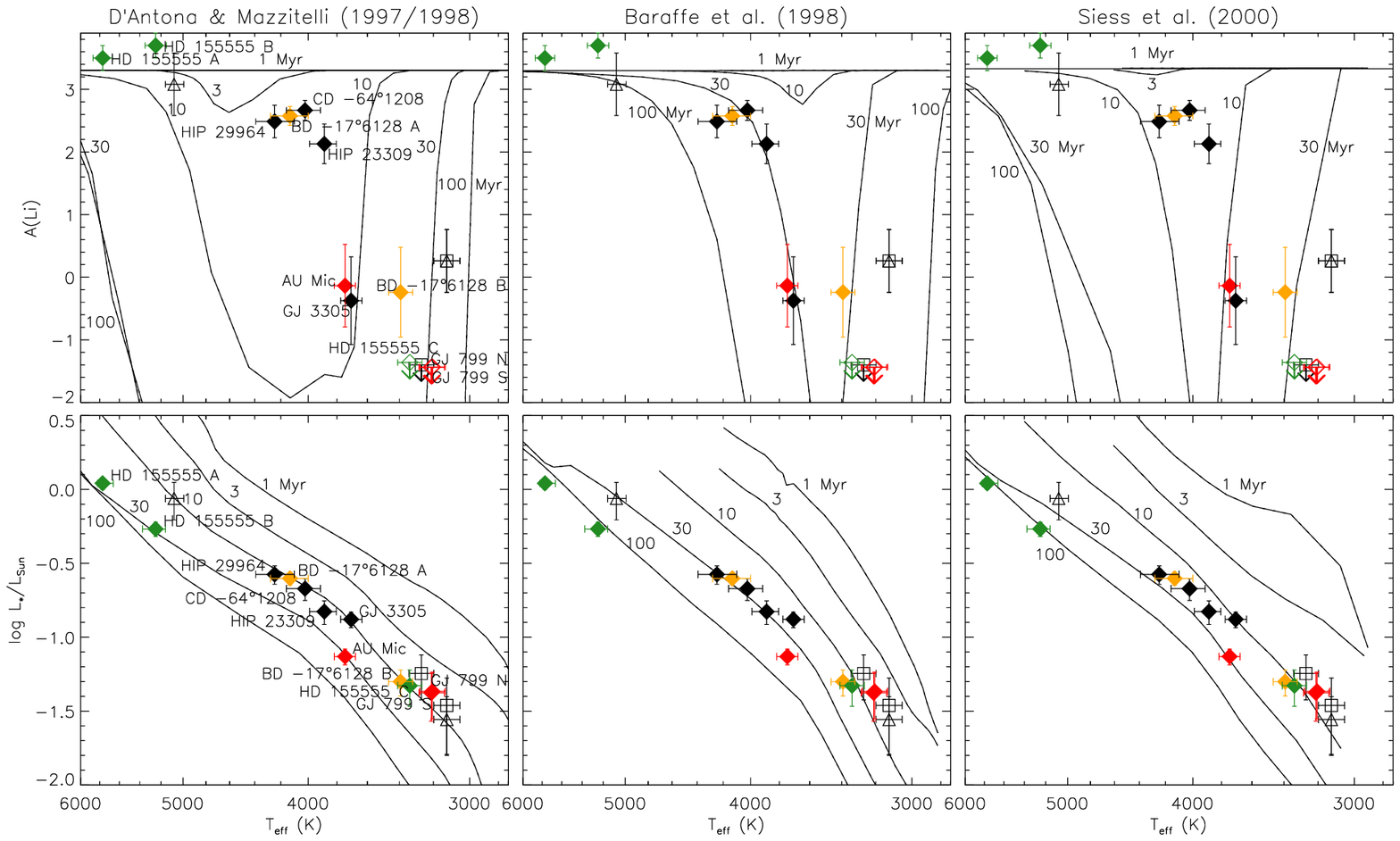}
\caption{Comparison of the data and theoretical models. The data are
  overlaid on three sets of models: \citet{Dantona97, Dantona98},
  \citet{Baraffe98}, and \citet{Siess00}. The results of this work are
  shown as diamonds. Single stars are shown in black and systems with
  multiple stars are shown in color (HD 155555 A/B/C (green); BD
  -17\arcdeg6128 A/B (orange); and AU Mic / GJ 799 N/S (red)). Upper
  limits are shown as open diamonds with arrows. The squares show HD
  112312 A/B \citep{Song02} and the triangles show V343 Nor A/B
  \citep{Torres06}. Note that the luminosities and effective
  temperatures for GJ 799 N/S are similar, so the two stars appear to
  be a single data point. Likewise, the lithium abundances and
  temperatures for HD 112312 B and V343 Nor B are the same. Notice
  that the ages predicted by the location on the HR diagram and by the
  location on the corresponding lithium abundance plot differ for any
  individual star. Taking the stars as a group, it is apparent that
  the models do not correctly predict the observed patterns of lithium
  depletion. This is also true of the stars in individual multiple systems
  whose components must be coeval. \label{fig:evolmod}}
\end{figure}

In Figure \ref{fig:evolmod} we plot our data against three sets of
evolutionary models: \citet{Dantona97,Dantona98}, \citet{Baraffe98},
and \citet{Siess00}. The \citet{Siess00} models include the effects of
convective overshooting. Known multiple systems are plotted in
color. We also plot HIP 112312 A/B and V343 Nor A/B, two systems that
cross the LDB. We calculated luminosities and effective temperatures
for these stars in the same manner as for our sample, using published
spectral types and V magnitudes
\citep{Zuckerman01,Song03,Torres06}. We used the lithium abundances
from \citet{Torres06}. Neither \citet{Song02} nor \citet{Torres06}
quote a lithium abundance for HIP 112312 A. Since the spectral type and
the upper limit in the equivalent width of this star given in
\citet{Song02} are similar to those of GJ 799 N, we assign HIP 1123112
A an upper limit of $A(\mathrm{Li}) = -1.4$ dex.

\begin{figure}
\includegraphics[width=6in]{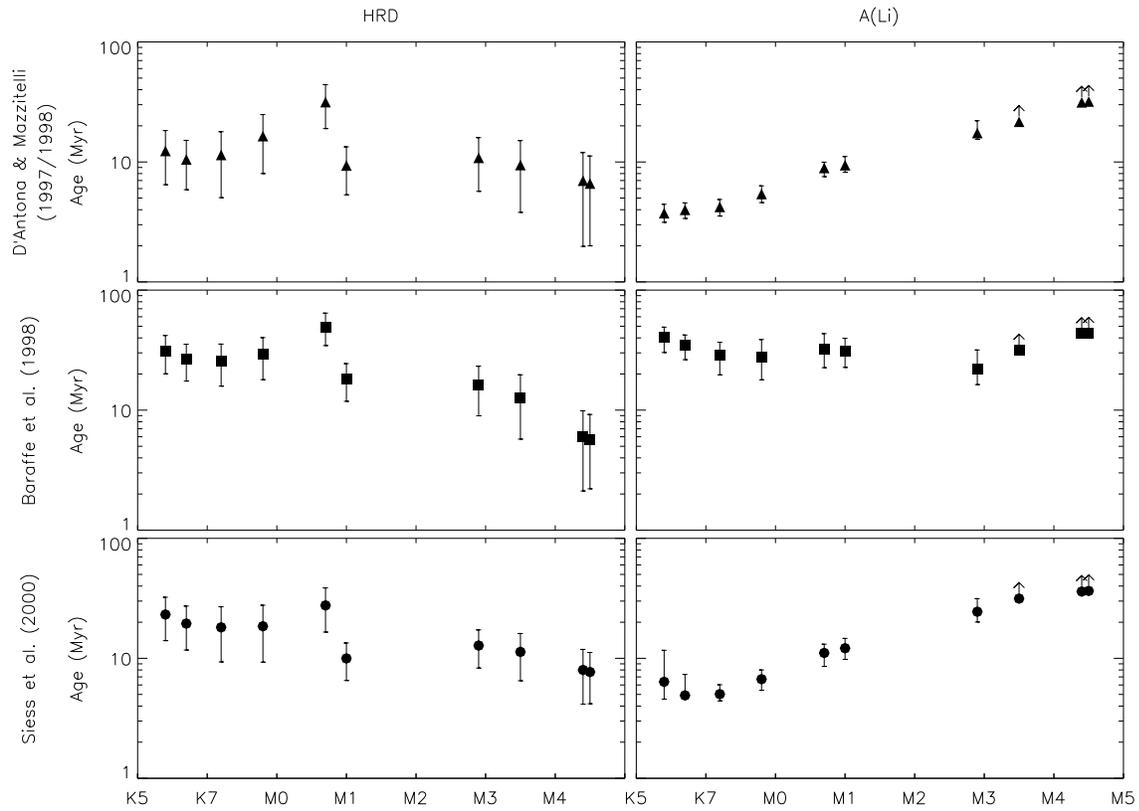}
\caption{Age as a function of spectral type, calculated for individual
stars from different models. The left column shows
ages calculated from the HR diagram and the right column shows ages
calculated from the lithium depletion isochrones. Since GJ 799 N and S
have the same spectral type, the points for GJ 799 S have been
displaced by $+0.1$ in spectral type for clarity.\label{fig:ages}}
\end{figure}

To explore the differences between the models and our data, we
independently determine ages for each star in two ways: from its
position on the HR diagram and from its position relative to the
lithium depletion isochrones. We estimate the uncertainty of the HR
diagram age by perturbing the effective temperatures and luminosities
within their uncertainties and adding the resulting age errors in
quadrature. Because the measured lithium abundance is correlated with
effective temperature, we estimate the uncertainties in the lithium
depletion age by simultaneously varying the effective temperatures and
lithium abundances within their uncertainties and taking the extreme
values for the age as a measure of its uncertainty. These data are
given in Table \ref{tab:ages} and shown in Figure \ref{fig:ages}. This
figure shows that the ages calculated from the HR diagram are roughly
consistent with a single age, but the ages measured from the lithium
depletion have systematic trends. The trend of increasing age with
spectral type found using the \citet{Dantona97,Dantona98} and
\citet{Siess00} lithium depletion isochrones simply reflects the fact
that the data cross the isochrones instead of lying parallel to them,
as seen in Figure \ref{fig:evolmod}.  Although the \citet{Baraffe98}
models are the closest to predicting a coeval group using lithium
depletion, they still fail to accurately represent the latest-type
stars in the sample. Even though the predicted ages are similar, the
pattern of lithium depletion seen in the \citet{Baraffe98} models is
the opposite of what is seen in our data. In particular, these lithium
depletion isochrones indicate that the lithium depletion boundary
occurs between 3400K and 3200K for the 30 Myr isochrone, which
approximates our data. Thus, we expect that stars at the cooler end of
the temperature range 3200-3400K should have a higher lithium
abundance than stars at the warmer end. In contrast, we observe that
HD 155555 C, GJ 799 N, and GJ 799 S with \teff $\sim$3300K, 3200K, and
3200K, respectively, have only upper limits in lithium abundance,
whereas the warmer star BD $-17^{\circ}6128$ B (\teff $\sim$ 3400K)
has a larger lithium abundance.

From these two methods for measuring ages, we observe two systematic
trends in the data. First, the ages of the mid-M dwarfs calculated
from the HR diagram isochrones are slightly younger than the ages of the
warmer stars. This effect has also been noted by
  \citet{Stassun2004} in the ages of binary components, and has been
  recognized more generally as a problem for PMS evolutionary models
  by \citet{Hillenbrand08}. The second trend is that the ages of
the mid-M stars calculated from their lithium depletion are older than the
ages of the warmer stars.

While the HR diagram isochrones match our data well and the stars are roughly
coeval, none of the evolutionary models have lithium depletion
isochrones consistent with a coeval group. If we consider each
multiple star system independently of the group as a whole, we find
similar inconsistencies in the predicted ages for the hotter and
cooler stars in each binary. Since the two stars in each binary are
almost certain to be coeval within $<1\,$ Myr\footnote{The
  multiplicity frequency of stars at the youngest observable
  pre--main-sequence ages is comparable to or greater than that among
  field stars \citep[][and references therein]{Duchene07}, indicating
  that binary formation is part of the star formation process.  Thus,
  we can consider binary systems to be $N=2$, coeval clusters to
  within $< 1$ Myr, although there are hints of smaller age
  differences \citep{Stassun2008}.}, this indicates that the mismatch
between our data and the isochrones is not the result of contamination
by unrelated stars or an extended period of star formation. In
general, the models seem to have difficulty describing the lithium
depletion of the lowest-mass stars.  In particular, the modeled rise
in lithium abundance at lower temperatures (the LDB) occurs at
temperatures too high for our stars in all of the evolutionary models.
Given the nominal HR diagram age, this means that an estimate of the age of
the group from the LDB will be systematically older than an age from
fitting the HR diagram regardless of the model used.

\begin{deluxetable}{lcc|cc|cc}
\tablewidth{0pt}
\tablecaption{Ages in Myr Calculated from Different Models and Isochrones\label{tab:ages}}
\tablehead{\colhead{ } & \multicolumn{2}{c|}{D'Antona \&}  & \multicolumn{2}{c|}{}\\
\colhead{} & \multicolumn{2}{c|}{Mazzitelli (1997, 1998)} & \multicolumn{2}{c|}{\citet{Baraffe98}}&\multicolumn{2}{c}{\citet{Siess00}}\\
\colhead{Star}&\colhead{HRD}& \multicolumn{1}{c|}{A(Li)}&\colhead{HRD}&\multicolumn{1}{c|}{A(Li)}&\colhead{HRD}&\colhead{A(Li)}\\}
\startdata
          HIP 29964 &
$12^{+ 6}_{- 6\phn}$&
$\phn4^{+ 1}_{- 1}$&
$31^{+11}_{-11}$&
$40^{+10}_{-13}$&
$23^{+ 9}_{- 9\phn}$&
$\phn6^{+ 6}_{- 2}$
\\
 BD $-17^{\circ}6128$ A &
$10^{+ 5}_{- 5\phn}$&
$\phn4^{+ 1}_{- 1}$&
$26^{+ 9}_{- 9\phn}$&
$35^{+ 9}_{- 9\phn}$&
$19^{+ 8}_{- 8\phn}$&
$\phn5^{+ 3}_{- 0}$
\\
         HIP 23309 &
$16^{+ 8}_{- 8\phn}$&
$\phn5^{+ 1}_{- 1}$&
$29^{+11}_{-11}$&
$28^{+12}_{-11}$&
$18^{+ 9}_{- 9\phn}$&
$\phn7^{+ 1}_{- 1}$
\\
   CD $-64^{\circ}1208$ &
$11^{+ 6}_{- 6\phn}$&
$\phn4^{+ 1}_{- 1}$&
$26^{+10}_{-10}$&
$29^{+10}_{-11}$&
$18^{+ 9}_{- 9\phn}$&
$\phn5^{+ 1}_{- 1}$
\\
            AU Mic &
$32^{+13}_{-13}$&
$\phn9^{+ 1}_{- 2}$&
$49^{+15}_{-15}$&
$32^{+12}_{-11}$&
$28^{+11}_{-11}$&
$11^{+ 2}_{- 3}$
\\
           GJ 3305 &
$\phn9^{+ 4}_{- 4\phn}$&
$\phn9^{+ 2}_{- 1}$&
$18^{+ 6}_{- 6\phn}$&
$31^{+10}_{- 9\phn}$&
$10^{+ 3}_{- 3\phn}$&
$12^{+ 3}_{- 2}$
\\
 BD $-17^{\circ}6128$ B &
$11^{+ 5}_{- 5\phn}$&
$17^{+ 5}_{- 2}$&
$16^{+ 7}_{- 7\phn}$&
$22^{+10}_{- 6\phn}$&
$13^{+ 4}_{- 4\phn}$&
$24^{+ 7}_{- 4}$
\\
       HD 155555 C &
$\phn9^{+ 6}_{- 6\phn}$&
$>22$&
$13^{+ 7}_{- 7\phn}$&
$>32$&
$11^{+ 5}_{- 5\phn}$&
$>31$
\\
          GJ 799 N &
$\phn7^{+ 5}_{- 5\phn}$&
$>31$&
$\phn6^{+ 4}_{- 4\phn}$&
$>43$&
$\phn8^{+ 4}_{- 4\phn}$&
$>36$
\\
          GJ 799 S &
$\phn7^{+ 5}_{- 5\phn}$&
$>32$&
$\phn6^{+ 4}_{- 4\phn}$&
$>44$&
$\phn8^{+ 4}_{- 4\phn}$&
$>36$
\\
\enddata
\end{deluxetable}

\section{Larger M-dwarf radii to the rescue?}
\label{section:radii}
\begin{figure}
\centering
\includegraphics{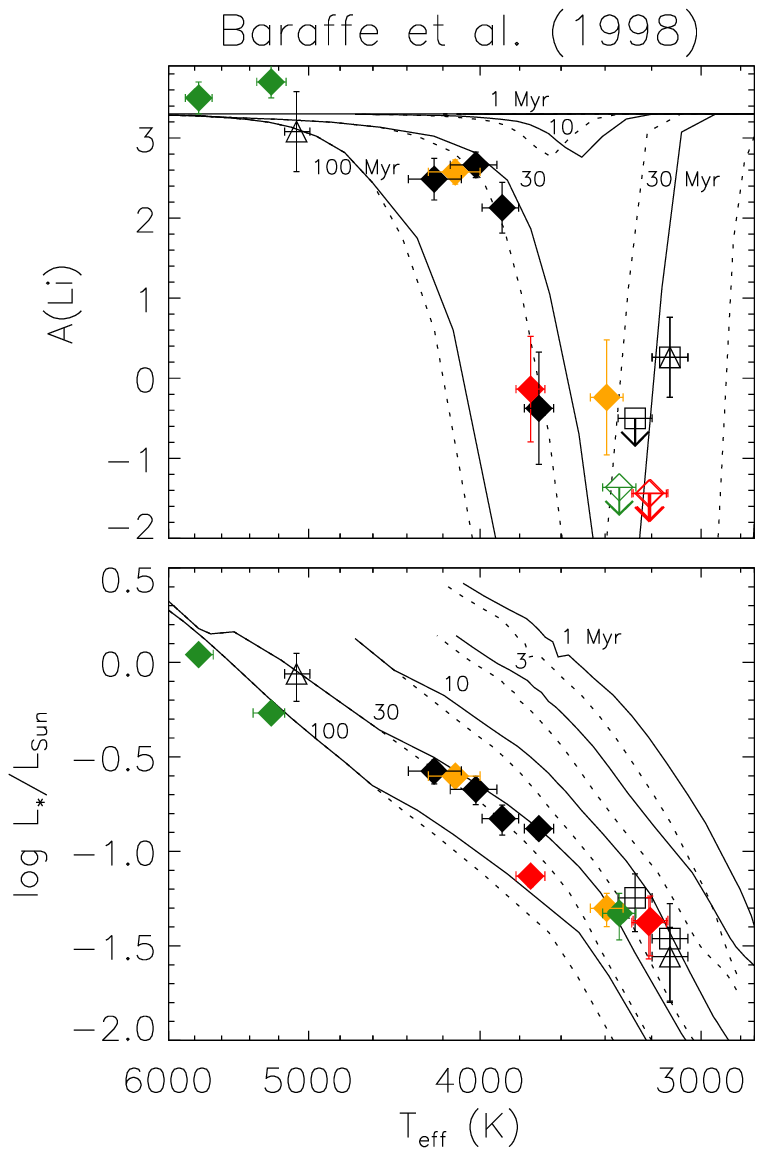}
\caption{\citet{Baraffe98} evolutionary models with artificially
inflated radii. The solid lines show the isochrones with an {\it ad hoc}
correction assuming that for \teff$\lesssim$4200K the radii should be
increased by 10\%. The dotted lines show the original model isochrones
from Fig. \ref{fig:evolmod}. The other symbols are the same as those
shown in Fig. \ref{fig:evolmod}.\label{fig:bcah}}
\end{figure}

We have shown that the pattern of Li depletion seen in the BPMG stars
is at odds with the predictions of PMS Li depletion models, in the
sense that the models under-predict the amount of Li depletion in the
M stars in our sample. The evolutionary models have a certain amount
of intrinsic uncertainty due to assumptions about mixing length,
equation of state, opacity, and atmosphere and boundary conditions,
whose effects on lithium depletion are discussed in some detail in
\citet{Burke04}.  Without a specific physical reason to change the
assumptions, it is difficult to decide which parameter or combination
of parameters should be changed in order to fit the lithium data.
However, there is an increasingly large body of observations (see
below) that evolutionary models tend to under-predict the radii of
low-mass stars, particularly M stars, which may point the way toward
the types of modifications to the models necessary to bring the HR
diagram ages and Li depletion ages into agreement.

Changes in stellar radii could affect the observed pattern of Li
depletion in two ways: by actually changing Li depletion as a function
of mass, and/or by altering the effective temperatures of stars, thus
shifting them relative to the model isochrones.  Regarding actual
changes in Li depletion, the work of \citet{King09} suggests a
connection between Li depletion and stellar radii. They propose that a
range in radii at a given stellar mass (perhaps due to a range in
rotation rates and/or chromospheric activity) would lead to a range of
interior temperature profiles and could explain the dispersion in
lithium depletion in the Pleiades.

However, even if the models are essentially correct in terms of the
interior conditions for a given mass, which largely govern the rate of
Li depletion, there could still be a disagreement between the model
predictions and the observations if the models do not correctly
predict the surface conditions, e.g.\ the radius or effective
temperature.  We suggest here that current data show evidence of
exactly this sort of disagreement.

There is mounting observational evidence that main-sequence stellar
evolutionary models predict radii for M dwarfs that are 10--20\%
smaller than the observed radii.  This effect is seen in stellar
angular diameters measured interferometrically
\citep{Lane2001,Segransan2003, Berger2006}, in radii determined from
eclipsing binaries \citep{Lopez-Morales2007, Ribas2008,
Fernandez2009}, and in radii determined from detailed modeling of
multi-band photometry to measure $L_{\rm Bol}$ and $T_{\rm eff}$
\citep{Mullan2001, Casagrande2008}.  There is still some question as
to whether this discrepancy exists only for stars that are the most
magnetically active \citep{Demory09}, or if it applies to all M dwarfs
\citep{Casagrande2008}.  There does seem to be at least some
differential effect from stellar activity \citep{Mullan2001,
Morales2008}, and models incorporating the effects of stellar activity
are better able to reproduce the observed radii \citep{Chabrier2007}.
For the present work, the question is moot as the stars in our sample
(and indeed all late-type pre--main-sequence stars) are very active,
with large X-ray luminosities and active chromospheres.

The discrepancy between the true radii and the radii predicted by
models very likely exists in the pre--main-sequence phase as well, if
it is due to effects of magnetic activity and/or missing opacity
sources in the models.  If this discrepancy is present, how would it
manifest itself? Here we argue that the effect would be similar to
what we observe, and correcting for this would bring HR diagram ages
and Li depletion ages closer together.  Corrected model isochrones for
a given age would need to move to lower $T_{\rm eff}$, increasing the
age inferred from the HR diagram and decreasing the age inferred from
Li depletion for a given star.

First consider the HR diagram.  Pre--main-sequence stars change
position in the HR diagram as they age precisely because their radii
are changing, as the stars descend (contract) toward the main
sequence.  If, at any given age, the true radii of stars are larger,
then the model isochrones for that age need to move up and to the
right.  Thus, the HR diagram age of a given star (of fixed, observed
\teff\,\ and \lbol) inferred from these corrected models would be
older than before.  Put another way, if radii are inflated by effects
other than youth, it will take longer for a star to contract to a
given radius, and thus to reach a given position on the HR diagram.

Now, consider the $A$(Li)--\teff\,\ plane.  The amount of Li depletion at
a given age is primarily driven by a star's mass, which sets its
interior conditions.  Inflating the radius of a star will cause that
star to have a cooler surface, i.e.\ a lower \teff.  Thus, corrected
model isochrones would need to shift to the right for a given $A$(Li)
and age.  Put another way, a star with an inflated radius will have a
hotter interior than we might otherwise expect given its surface
appearance, and thus will exhibit more Li depletion at a given age and
\teff\,\ than would be predicted if the radius were not modified.
This matches well with the enhanced Li depletion we observe for the
cooler stars.

Calculating new pre--main-sequence evolutionary models incorporating
these effects is beyond the scope of this work, but we can explore
these effects to a limited degree by introducing an {\it ad hoc} shift
to the existing model isochrones.  As a test of this idea, we shifted
the model isochrones of \citet{Baraffe98} and compared them to our
observations of A(Li) and \teff\,\ for the BPMG stars.  Based on the
observations of M stars discussed above, we assumed that the larger
radii are present only for stars cooler than $\sim$ 4200 K
\citep{Casagrande2008}, and that the model radii should be inflated by
10\%.  As shown in Figure \ref{fig:bcah}, these corrections to the
models do indeed bring the inferred Li depletion and HR diagram ages
of the cooler stars into better agreement, though some discrepancies
remain.  Clearly, a more systematic treatment of this effect is
needed, perhaps taking into account effects of different activity
levels as well as different masses.

To close, we note that one implication of this result is that neither
the LDB ages nor the HR diagram ages of M stars are correct, but that
the answer lies somewhere between the two.

\section{Conclusions}
\label{sec:conclusions}
We have measured lithium abundances and effective temperatures for ten
members of the $\beta$ Pictoris Moving Group. We compare these
abundances to the predictions of the evolutionary models of
\citet{Dantona97,Dantona98}, \citet{Baraffe98}, and \citet{Siess00}.
We find that while the HR diagram isochrones of these models reproduce
our data fairly well, the lithium depletion isochrones do not. In
particular, the models predict less lithium depletion for the
latest-type stars than we observe. Thus, the ages determined based on
the lithium abundances are older than the ages determined from the HR
diagram.

The most striking result is the non-detection of lithium in the
latest-type members (M3.5--M5) of this group. The lack of an observable
lithium line in HD 155555 C, GJ 799 N, and GJ 799 S indicates that
lithium depletion proceeds at a much faster rate than is predicted in
the theoretical models given their spectral types. One potential clue
lies in the discrepancy between the observed radii of M dwarfs and the
radii predicted by models.

Regardless of the physical mechanism, it appears that mid-M stars
deplete their lithium rapidly. The presence of the lithium line at
6707.8\AA\, in a star of this type remains a clear indicator of youth
and suggests an age younger than that of the $\beta$ Pictoris Moving
Group, while the lack of a detectable line is {\it not} a strong
indicator that a star is old.

\acknowledgements{We thank the referee for useful comments that
  improved this work.  We thank Marc Pinsonneault, Keivan Stassun, and
  Peter Hauschildt for helpful conversations. We also thank Christine
  Johnas and Blair Reaser for early input into this work.  EJ
  gratefully acknowledges the support of a Eugene M. Lang Faculty
  Fellowship from Swarthmore College, and NSF grant AST-0307830.}

\bibliographystyle{/home/morgan/jyee/tex/apj}

\begin{thebibliography}
\expandafter\ifx\csname natexlab\endcsname\relax\def\natexlab#1{#1}\fi

\bibitem[{{Baraffe} {et~al.}(1998){Baraffe}, {Chabrier}, {Allard}, \&
  {Hauschildt}}]{Baraffe98}
{Baraffe}, I., {Chabrier}, G., {Allard}, F., \& {Hauschildt}, P.~H. 1998, \aap,
  337, 403

\bibitem[{{Barrado y Navascu{\'e}s} {et~al.}(2004){Barrado y Navascu{\'e}s},
  {Stauffer}, \& {Jayawardhana}}]{Barrado04}
{Barrado y Navascu{\'e}s}, D., {Stauffer}, J.~R., \& {Jayawardhana}, R. 2004,
  \apj, 614, 386

\bibitem[Barrado y Navascu{\'e}s, Stauffer, \&
Patten(1999)]{Barrado99} Barrado y Navascu{\'e}s, D., Stauffer, J.~R.,
\& Patten, B.~M.\ 1999, \apjl, 522, L53

\bibitem[{{Basri} {et~al.}(1996){Basri}, {Marcy}, \& {Graham}}]{Basri96}
{Basri}, G., {Marcy}, G.~W., \& {Graham}, J.~R. 1996, \apj, 458, 600

\bibitem[{{Berger} {et~al.}(2006){Berger}, {Gies}, {McAlister}, {ten
  Brummelaar}, {Henry}, {Sturmann}, {Sturmann}, {Turner}, {Ridgway},
  {Aufdenberg}, \& {M{\'e}rand}}]{Berger2006}
{Berger}, D.~H., {Gies}, D.~R., {McAlister}, H.~A., {ten Brummelaar}, T.~A.,
  {Henry}, T.~J., {Sturmann}, J., {Sturmann}, L., {Turner}, N.~H., {Ridgway},
  S.~T., {Aufdenberg}, J.~P., \& {M{\'e}rand}, A. 2006, \apj, 644, 475

\bibitem[{{Bildsten} {et~al.}(1997){Bildsten}, {Brown}, {Matzner}, \&
  {Ushomirsky}}]{Bildsten97}
{Bildsten}, L., {Brown}, E.~F., {Matzner}, C.~D., \& {Ushomirsky}, G. 1997,
  \apj, 482, 442

\bibitem[{{Burke} {et~al.}(2004){Burke}, {Pinsonneault}, \& {Sills}}]{Burke04}
{Burke}, C.~J., {Pinsonneault}, M.~H., \& {Sills}, A. 2004, \apj, 604, 272

\bibitem[{{Casagrande} {et~al.}(2008){Casagrande}, {Flynn}, \&
  {Bessell}}]{Casagrande2008}
{Casagrande}, L., {Flynn}, C., \& {Bessell}, M. 2008, \mnras, 389, 585

\bibitem[{{Chabrier} {et~al.}(2007){Chabrier}, {Gallardo}, \&
  {Baraffe}}]{Chabrier2007}
{Chabrier}, G., {Gallardo}, J., \& {Baraffe}, I. 2007, \aap, 472, L17

\bibitem[{{D'Antona} \& {Mazzitelli}(1997)}]{Dantona97}
{D'Antona}, F. \& {Mazzitelli}, I. 1997, Memorie della Societa Astronomica
  Italiana, 68, 807

\bibitem[{{D'Antona} \& {Mazzitelli}(1998)}]{Dantona98}
{D'Antona}, F. \& {Mazzitelli}, I. 1998, in Astronomical Society of the Pacific
  Conference Series, Vol. 134, Brown Dwarfs and Extrasolar Planets, ed.
  R.~{Rebolo}, E.~L. {Martin}, \& M.~R. {Zapatero Osorio}, 442--+

\bibitem[{{Demory} {et~al.}(2009){Demory}, {Segransan}, {Forveille}, {Queloz},
  {Beuzit}, {Delfosse}, {Di Folco}, {Kervella}, {Le Bouquin}, \&
  {Perrier}}]{Demory09}
{Demory}, B., {Segransan}, D., {Forveille}, T., {Queloz}, D., {Beuzit}, J.,
  {Delfosse}, X., {Di Folco}, E., {Kervella}, P., {Le Bouquin}, J., \&
  {Perrier}, C. 2009, ArXiv e-prints

\bibitem[Duch{\^e}ne et al.(2007)]{Duchene07} Duch{\^e}ne, G.,
Delgado-Donate, E., Haisch, K.~E., Jr., Loinard, L., \&
Rodr{\'{\i}}guez, L.~F.\ 2007, Protostars and Planets V, 379

\bibitem[{{Fernandez} {et~al.}(2009){Fernandez}, {Latham}, {Torres}, {Everett},
  {Mandushev}, {Charbonneau}, {O'Donovan}, {Alonso}, {Esquerdo},
  {Hergenrother}, \& {Stefanik}}]{Fernandez2009}
{Fernandez}, J.~M., {Latham}, D.~W., {Torres}, G., {Everett}, M.~E.,
  {Mandushev}, G., {Charbonneau}, D., {O'Donovan}, F.~T., {Alonso}, R.,
  {Esquerdo}, G.~A., {Hergenrother}, C.~W., \& {Stefanik}, R.~P. 2009, \apj,
  701, 764

\bibitem[{{Fernandez-Figueroa} {et~al.}(1993){Fernandez-Figueroa}, {Barrado},
  {de Castro}, \& {Cornide}}]{Fernandez93}
{Fernandez-Figueroa}, M.~J., {Barrado}, D., {de Castro}, E., \& {Cornide}, M.
  1993, \aap, 274, 373

\bibitem[{{Hillenbrand} {et~al.}(2008){Hillenbrand}, {Bauermeister}, \&
  {White}}]{Hillenbrand08}
{Hillenbrand}, L.~A., {Bauermeister}, A., \& {White}, R.~J. 2008, in
  Astronomical Society of the Pacific Conference Series, Vol. 384, 14th
  Cambridge Workshop on Cool Stars, Stellar Systems, and the Sun, ed. G.~{van
  Belle}, 200--+

\bibitem[{{Jeffries}(2000)}]{Jeffries00}
{Jeffries}, R.~D. 2000, in Astronomical Society of the Pacific Conference
  Series, Vol. 198, Stellar Clusters and Associations: Convection, Rotation,
  and Dynamos, ed. R.~{Pallavicini}, G.~{Micela}, \& S.~{Sciortino}, 245--+

\bibitem[{{Jeffries} \& {Naylor}(2001)}]{Jeffries01}
{Jeffries}, R.~D. \& {Naylor}, T. 2001, in Astronomical Society of the Pacific
  Conference Series, Vol. 243, From Darkness to Light: Origin and Evolution of
  Young Stellar Clusters, ed. T.~{Montmerle} \& P.~{Andr{\'e}}, 633--+

\bibitem[{{Jeffries} \& {Oliveira}(2005)}]{Jeffries05}
{Jeffries}, R.~D. \& {Oliveira}, J.~M. 2005, \mnras, 358, 13

\bibitem[{{Kalas} {et~al.}(2004){Kalas}, {Liu}, \& {Matthews}}]{Kalas04}
{Kalas}, P., {Liu}, M.~C., \& {Matthews}, B.~C. 2004, Science, 303, 1990

\bibitem[{{Kenyon} \& {Hartmann}(1995)}]{Kenyon95}
{Kenyon}, S.~J. \& {Hartmann}, L. 1995, \apjs, 101, 117

\bibitem[{{King} {et~al.}(2009){King}, {Jeremy}}]{King09}
{King}, J.~R. et al. 2009, \apj, Submitted

\bibitem[{{Kirkpatrick} {et~al.}(1991){Kirkpatrick}, {Henry}, \&
  {McCarthy}}]{Kirkpatrick91}
{Kirkpatrick}, J.~D., {Henry}, T.~J., \& {McCarthy}, Jr., D.~W. 1991, \apjs,
  77, 417

\bibitem[Lane et al.(2001)]{Lane2001} Lane, B.~F., Boden, A.~F., \&
Kulkarni, S.~R.\ 2001, \apjl, 551, L81

\bibitem[{{L{\'o}pez-Morales}(2007)}]{Lopez-Morales2007}
{L{\'o}pez-Morales}, M. 2007, \apj, 660, 732

\bibitem[{{Luhman}(1999)}]{Luhman99}
{Luhman}, K.~L. 1999, \apj, 525, 466

\bibitem[Manzi et al.(2008)]{Manzi08} Manzi, S., Randich, S., de Wit,
W.~J., \& Palla, F.\ 2008, \aap, 479, 141

\bibitem[{{Mentuch} {et~al.}(2008){Mentuch}, {Brandeker}, {van Kerkwijk},
  {Jayawardhana}, \& {Hauschildt}}]{Mentuch08}
{Mentuch}, E., {Brandeker}, A., {van Kerkwijk}, M.~H., {Jayawardhana}, R., \&
  {Hauschildt}, P.~H. 2008, \apj, 689, 1127

\bibitem[{{Morales} {et~al.}(2008){Morales}, {Ribas}, \& {Jordi}}]{Morales2008}
{Morales}, J.~C., {Ribas}, I., \& {Jordi}, C. 2008, \aap, 478, 507

\bibitem[{{Mullan} \& {MacDonald}(2001)}]{Mullan2001}
{Mullan}, D.~J. \& {MacDonald}, J. 2001, \apj, 559, 353

\bibitem[{{Neuh{\"a}user} {et~al.}(2002){Neuh{\"a}user}, {Guenther},
  {Mugrauer}, {Ott}, \& {Eckart}}]{Neuhauser02}
{Neuh{\"a}user}, R., {Guenther}, E., {Mugrauer}, M., {Ott}, T., \& {Eckart}, A.
  2002, \aap, 395, 877

\bibitem[{{Neuh{\"a}user} (2009)}]{Neuhauser09}
{Neuh{\"a}user},~R., 2009, private communication

\bibitem[{{Palla} {et~al.}(2007){Palla}, {Randich}, {Pavlenko}, {Flaccomio}, \&
  {Pallavicini}}]{Palla07}
{Palla}, F., {Randich}, S., {Pavlenko}, Y.~V., {Flaccomio}, E., \&
  {Pallavicini}, R. 2007, \apjl, 659, L41

\bibitem[{{Pallavicini} {et~al.}(1992){Pallavicini}, {Randich}, \&
  {Giampapa}}]{Pallavicini92}
{Pallavicini}, R., {Randich}, S., \& {Giampapa}, M.~S. 1992, \aap, 253, 185

\bibitem[{{Pasquini} {et~al.}(1991){Pasquini}, {Cutispoto}, {Gratton}, \&
  {Mayor}}]{Pasquini91}
{Pasquini}, L., {Cutispoto}, G., {Gratton}, R., \& {Mayor}, M. 1991, \aap, 248,
  72

\bibitem[{{Randich} {et~al.}(1993){Randich}, {Gratton}, \&
  {Pallavicini}}]{Randich93}
{Randich}, S., {Gratton}, R., \& {Pallavicini}, R. 1993, \aap, 273, 194

\bibitem[{{Reid} {et~al.}(1995){Reid}, {Hawley}, \& {Gizis}}]{Reid95}
{Reid}, I.~N., {Hawley}, S.~L., \& {Gizis}, J.~E. 1995, \aj, 110, 1838

\bibitem[{{Riaz} {et~al.}(2006){Riaz}, {Gizis}, \& {Harvin}}]{Riaz06}
{Riaz}, B., {Gizis}, J.~E., \& {Harvin}, J. 2006, \aj, 132, 866

\bibitem[{{Ribas} {et~al.}(2008){Ribas}, {Morales}, {Jordi}, {Baraffe},
  {Chabrier}, \& {Gallardo}}]{Ribas2008}
{Ribas}, I., {Morales}, J.~C., {Jordi}, C., {Baraffe}, I., {Chabrier}, G., \&
  {Gallardo}, J. 2008, Memorie della Societa Astronomica Italiana, 79, 562

\bibitem[{{Scholz} {et~al.}(2007){Scholz}, {Coffey}, {Brandeker}, \&
  {Jayawardhana}}]{Scholz07}
{Scholz}, A., {Coffey}, J., {Brandeker}, A., \& {Jayawardhana}, R. 2007, \apj,
  662, 1254

\bibitem[S{\'e}gransan et al.(2003)]{Segransan2003} S{\'e}gransan, D.,
Kervella, P., Forveille, T., \& Queloz, D.\ 2003, \aap, 397, L5

\bibitem[{{Sestito} {et~al.}(2008){Sestito}, {Palla}, \& {Randich}}]{Sestito08}
{Sestito}, P., {Palla}, F., \& {Randich}, S. 2008, \aap, 487, 965

\bibitem[{{Siess} {et~al.}(2000){Siess}, {Dufour}, \& {Forestini}}]{Siess00}
{Siess}, L., {Dufour}, E., \& {Forestini}, M. 2000, \aap, 358, 593

\bibitem[{{Soderblom} {et~al.}(1993{\natexlab{a}}){Soderblom}, {Fedele},
  {Jones}, {Stauffer}, \& {Prosser}}]{Soderblom93iv}
{Soderblom}, D.~R., {Fedele}, S.~B., {Jones}, B.~F., {Stauffer}, J.~R., \&
  {Prosser}, C.~F. 1993{\natexlab{a}}, \aj, 106, 1080

\bibitem[{{Soderblom} {et~al.}(1993{\natexlab{b}}){Soderblom}, {Jones},
  {Balachandran}, {Stauffer}, {Duncan}, {Fedele}, \& {Hudon}}]{Soderblom93iii}
{Soderblom}, D.~R., {Jones}, B.~F., {Balachandran}, S., {Stauffer}, J.~R.,
  {Duncan}, D.~K., {Fedele}, S.~B., \& {Hudon}, J.~D. 1993{\natexlab{b}}, \aj,
  106, 1059

\bibitem[{{Song} {et~al.}(2002){Song}, {Bessell}, \& {Zuckerman}}]{Song02}
{Song}, I., {Bessell}, M.~S., \& {Zuckerman}, B. 2002, \apjl, 581, L43

\bibitem[{{Song} {et~al.}(2003){Song}, {Zuckerman}, \& {Bessell}}]{Song03}
{Song}, I., {Zuckerman}, B., \& {Bessell}, M.~S. 2003, \apj, 599, 342

\bibitem[Stassun et al.(2004)]{Stassun2004} Stassun, K.~G., Mathieu,
R.~D., Vaz, L.~P.~R., Stroud, N., \& Vrba, F.~J.\ 2004, \apjs, 151,
357

\bibitem[Stassun et al.(2008)]{Stassun2008} Stassun, K.~G., Mathieu,
R.~D., Cargile, P.~A., Aarnio, A.~N., Stempels, E., \& Geller, A.\
2008, \nat, 453, 1079

\bibitem[Stauffer et al.(1999)]{Stauffer99} Stauffer, J.~R., et al.\
1999, \apj, 527, 219

\bibitem[{{Torres} {et~al.}(2006){Torres}, {Quast}, {da Silva}, {de La Reza},
  {Melo}, \& {Sterzik}}]{Torres06}
{Torres}, C.~A.~O., {Quast}, G.~R., {da Silva}, L., {de La Reza}, R., {Melo},
  C.~H.~F., \& {Sterzik}, M. 2006, \aap, 460, 695

\bibitem[{{van Leeuwen}(2007)}]{vanLeeuwen07}
{van Leeuwen}, F., ed. 2007, Astrophysics and Space Science Library, Vol. 350,
  {Hipparcos, the New Reduction of the Raw Data}

\bibitem[{{White} \& {Hillenbrand}(2005)}]{White05}
{White}, R.~J. \& {Hillenbrand}, L.~A. 2005, \apjl, 621, L65

\bibitem[{{Zuckerman} \& {Song}(2004)}]{Zuckerman04}
{Zuckerman}, B. \& {Song}, I. 2004, \araa, 42, 685

\bibitem[{{Zuckerman} {et~al.}(2001){Zuckerman}, {Song}, {Bessell}, \&
  {Webb}}]{Zuckerman01}
{Zuckerman}, B., {Song}, I., {Bessell}, M.~S., \& {Webb}, R.~A. 2001, \apjl,
  562, L87

\end{thebibliography}

\end{document}